\documentstyle[prl,aps,multicol,epsf]{revtex}
\newcommand{\Sla}[1]{{\not \! \! {#1}}}
\newcommand{\sla}[1]{{\not \! \! \: {#1}}}
\newcommand{\slaD}{\sla{D}}

\newcommand{\slad}{\Sla{\, d}}
\begin{document}
\input epsf 
\title{Axial Anomaly in Quasi-1D Chiral Superfluids}
\draft
\author{J. Goryo} 
\address{\it Department of Physics, Hokkaido University, 
Sapporo, 060-0810 Japan}
\date{\today}
\maketitle

\begin{abstract}
The axial anomaly in a quasi-one-dimensional (quasi-1D) 
chiral $p$-wave superfluid model, which has a 
$\varepsilon_{x} p_{x}+i \varepsilon_{y} p_{y}$-wave gap in 2D is studied.  
The anomaly causes an accumulation of the quasiparticle 
and a quantized chiral current density in an inhomogeneous magnetic field. 
These effects are related to the winding number of the gap.  
By varying the parameters $\varepsilon_{x}$ and $\varepsilon_{y}$,  
the model could be applicable to Sr$_{2}$RuO$_{4}$ near the second 
superconducting transition point, some quasi-1D organic superconductors
and the fractional quantum Hall state at $\nu = 5 / 2$ 
Landau level filling factor. 

\end{abstract}


\begin{multicols}{2}


The chiral superfluidity is realized 
in the superfluid $^{3}$He-A \cite{volovik-1}. 
Recently, the possibility of 
the chiral superconductivity is argued\cite{cpsc,cdsc}. 
In such superfluids or superconductors, the ground state is the condensate of 
the Cooper pairs which have orbital angular 
momentum along a same direction. Therefore, time-reversal symmetry (T) and 
also parity (P) in two-dimensional space (2D) are violating. 
We investigate a quasi-1D chiral $p$-wave superfluid in 2D.    
It is revealed that the axial anomaly causes 
P- and T-violating phenomena related to the quantized number.   

The axial anomaly has been originally pointed out 
in the Dirac QED in 3D\cite{a-b,b-j}. It is a phenomenon 
that a symmetry under the phase transformation $e^{i \gamma_{5} \alpha}$ 
of the Dirac field in the action 
at the classical level is broken in the quantum theory. 
Here, $\alpha$ is a constant and $\gamma_{5}$ is a hermitian 
matrix which anti-commutes with all of the Dirac matrices $\gamma_{\mu}$, 
where $\mu$ is the spacetime index. 
The Adler-Bardeen's theorem guarantees the absence of higher order 
collections to the divergence of the axial current\cite{a-b}. 
Therefore, the exact calculation of the two-photon decay rate 
of neutral $\pi$ meson can be done.
It has been pointed out that the same results are 
obtained by using the path-integral formalism and has been clarified 
the relation between the axial anomaly and topological 
quantized numbers through the Atiyah-Singer index theorem\cite{fujikawa}. 

It has been pointed out that the axial anomaly also plays important role  
in the quantum Hall effect (QHE) in the 2D massive Dirac QED.   
In 2D, the mass term of the Dirac Fermion violates P and T like 
the magnetic field, and 
the Hall effect may occur. 
It was shown that the existence of the Hall current and its quantization  
are caused by the axial anomaly in 1D\cite{red-nie-sem,ishikawa}.       
The relation between the axial anomaly and QHE     
in 2D electron gas in the magnetic field was 
also discussed\cite{ishikawa,edge}, and the quantized Hall conductance is 
expressed by the winding number of the fermionic propagator in 
the momentum space\cite{i-i-m-t}.  
Other applications of the axial anomaly 
to the condensed matter physics are studied in 
the field of the superfluid $^{3}$He in 3D 
and in charge density waves in 1D conductors\cite{volovik-1,he-3-cdw-anomaly}. 

The phenomena caused by the axial anomaly are  
related to the topologically quantized numbers.  
On the analogy of QHE, it is expected that the axial anomaly 
also plays important role in other P- and T-violating 2D systems. 
 In this letter, we investigate the axial anomaly in a quasi-1D 
chiral superfluid model in 2D, which has the spin-triplet 
$\varepsilon_{x} p_{x} + i \varepsilon_{y} p_{y}$-wave symmetry. 
P and T-violation occur whenever 
both of $\varepsilon_{x}$ and $\varepsilon_{y}$ are non-zero. 
We show that the axial anomaly in 1D causes an accumulation of    
the mass density of the quasiparticle in an inhomogeneous magnetic field.  
The axial anomaly also causes a chiral current density, which is 
perpendicular to the gradient of the magnetic field.  
These effects are related to the winding number of   
the gap; ${\rm sgn}(\varepsilon_{x} \varepsilon_{y})$\cite{vol-g-i}. 
Our discussion would be valid for the superconductors  
by taking into account the Meissner effect. 
By varying the parameters $\varepsilon_{x}$ and $\varepsilon_{y}$,  
the model could be applicable to  
Sr$_{2}$RuO$_{4}$ near the second superconducting transiton point\cite{s-r-j}, 
and some quasi-1D organic superconductors\cite{orgsc,lee,kohmoto}
or the fractional quantum Hall (FQH) state at $\nu = 5 / 2$ 
Landau level (LL) filling factor \cite{maeda,5/2exp,pfaff}.   
We use the 2+1-dimensional Euclidian spacetime and 
the natural unit ($\hbar=c=1$) in the present paper.

Let us consider a quasi-1D chiral superfluid model.   
We assume a linearized fermion spectrum and 
a spin-triplet chiral $p$-wave gap near the Fermi surface 
in the normal state written as,  
\begin{eqnarray}
\epsilon_{\rm R,L}({\bf p})&=&\pm v_{\rm F} (p_{x} \mp p_{\rm F}), 
\label{udkin}\\ 
\Delta({\bf p}) &=& i \sigma_{3} \sigma_{2} 
\frac{\Delta}{|p_{F}|}(\varepsilon_{x} p_{x} +  i \varepsilon_{y} p_{y}),
\label{c-p-w}
\end{eqnarray} 
where 
$v_{\rm F}$ and $p_{\rm F}$ are the Fermi velocity and the Fermi momentum, 
respectively. $\epsilon_{R} ({\bf p})$ ($\epsilon_{L} ({\bf p})$) is 
the kinetic energy for the right (left) mover. 
When $\varepsilon_{x}<<1$ and $\varepsilon_{y}\sim 1$, 
the model describes the low energy excitations 
(the quasiparticle excitations around the tiny gap points)  
of Sr$_{2}$RuO$_{4}$ near the second superconducting transition point 
under the uniaxial pressure in the $x-y$ plane (the 
basal plane)\cite{s-r-j}. For simplicity, 
we assume a circular Fermi Surface in the normal state (Fig.1(a)).  
When $\varepsilon_{x}\sim1$ and $\varepsilon_{y}<< 1$, 
the model describes the excitations near $p_{x}=\pm p_{\rm F}$  
with the chiral $p$-wave gap, whose 
kinetic energy in the normal state is  
$\epsilon({\bf p})= - 2 t_{x} \cos(p_{x} a) - 2 t_{y} \sin(p_{y} b) 
- \epsilon_{\rm F}$ $(t_{y}<<t_{x}, \epsilon_{\rm F}$; the Fermi energy). 
The model in this case is applicable to 
some quasi-1D organic superconductors 
or the FQH state at $\nu = 5 / 2$ LL   
filling factor (Fig.1(b)).    
The quasi-1D superconductivity  
has been observed in organic conductors, such as (TMTSF)$_{2}$X\cite{orgsc}.  
The NMR knight shift study in Ref.\cite{lee} is a evidence supporting 
a spin-triplet pairing state in (TMTSF)$_{2}$PF$_{6}$. 
The spin-triplet superconductivity in a quasi-1D system 
with a nodeless gap is obtained theoretically 
when an electron-phonon coupling and 
antiferromagnetic fluctuations are taken into account\cite{kohmoto}.
Our discussion would be valid for such superconductors if they have 
the chiral $p$-wave pairing symmetry.
It has been pointed out that the unidirectional charge density wave state, 
which has the belt-shaped Fermi sea like Fig. 1(b), seems to be 
the most plausible compressible state at the half-filled Landau levels 
in the quantum Hall system\cite{maeda}. 
Recently, the FQH effect has been observed at $\nu=5 / 2$ 
\cite{5/2exp}, and the effect could be described by  
the chiral $p$-wave pairing state (the Pfaffian state)\cite{pfaff}.   
Therefore, our model could be a candidate of 
the $\nu=5 / 2$ FQH state.

\vspace{0cm}
\begin{figure}
\centerline{
\epsfysize=7cm\epsffile{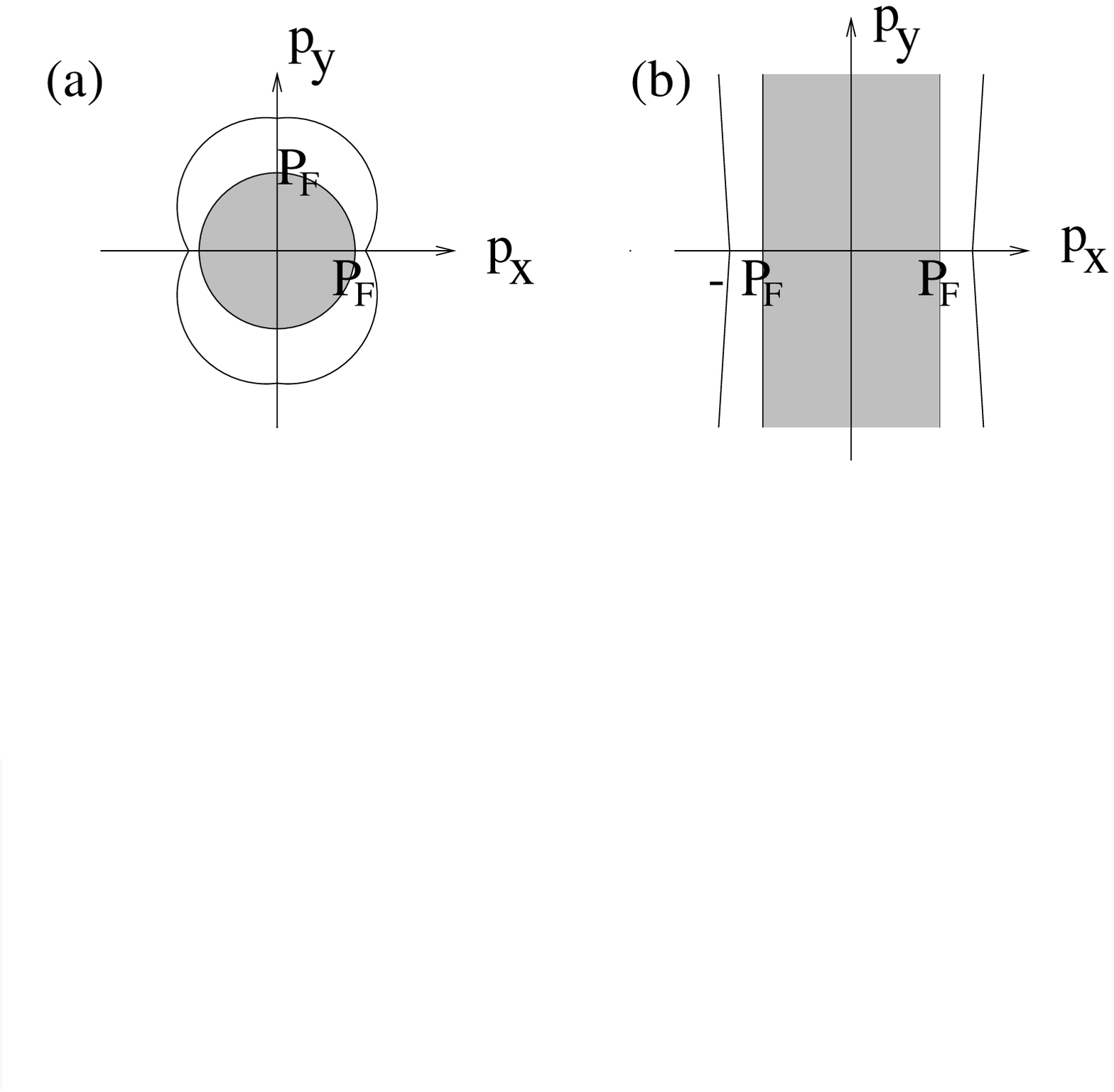}}
\vspace{-4cm}
Fig. 1. The Fermi sea in the normal state and 
the momentum dependence of the gap functions 
for (a) Sr$_{2}$RuO$_{4}$ near the second superconducting 
phase transition point, which corresponds to   
$\varepsilon_{x} <<1$, $\varepsilon_{y}\sim 1$ in our model,   
and for (b) some quasi-1D organic superconductors with 
the chiral $p$-wave pairing symmetry or 
the FQH state at $\nu=5 / 2$ LL
filling factor, which corresponds to 
$\varepsilon_{x} \sim 1$, $\varepsilon_{y}<<1$. 
The shadows show the Fermi sea, and the distance between 
outer lines and inner lines shows the magnitude of the gap. 
\end{figure}

The Lagrangian of our model is written as 
\begin{eqnarray} 
{\cal {L}}&=&\bar{\Psi}_{\bf p} 
\left[\{ i \partial_{\tau} + \mu \frac{d {\rm B}_{z}}{d y} y \sigma_{3} \} 
\otimes \gamma_{\tau} \right.   
\label{ucpsc-lag}\\
&&\left.+ \frac{\Delta}{|p_{\rm F}|} \sigma_{3} \otimes
( \varepsilon_{x} p_{x} \gamma_{x} + \varepsilon_{y} p_{y} \gamma_{y} ) 
- i v_{\rm F} (p_{x} - p_{\rm F}) \right] \Psi_{\bf p}.       
\nonumber
\end{eqnarray} 
Here, we use the Bogoliubov-Nambu representation with an isospin $\alpha=1,2$
$$
\Psi({\bf x})=e^{i p_{\rm F} x} 
\left(\begin{array}{c}
\psi(p_{\rm F},{\bf x})  \\ 
i \sigma_{2} \psi^{*}(-p_{\rm F},{\bf x})  
\end{array} \right),  
$$ 
and $\Psi_{\bf p}$ is its Fourier transform. 
$\psi(p_{\rm F},{\bf x})$ and $\psi(-p_{\rm F},{\bf x})$ are
the slowly varying fields for the right mover and the left mover 
with a real spin index $s=1,2$, 
respectively.  The matrices 
$\gamma_{\tau}=\tau_{3}, \gamma_{x}=-\tau_{2}$ and $\gamma_{y}=-\tau_{1}$ 
are the $2\times2$ Pauli matrices with isospin indices and 
$\sigma_{i}~(i=1,2,3)$ is the $2\times2$ Pauli matrices with spin indices.  
The symbol $\sigma_{i} \otimes \gamma_{\tau,x,y}$ shows the direct product.
$\bar{\Psi}$ is defined as $\bar{\Psi}= - i \Psi^{\dagger} \gamma_{\tau}$.   
We assume a magnetic field, which is directed to the $z$-axis 
(the $c$-axis in the crystal) and  
has a constant gradient in the $y$-direction, 
i.e. $B_{z}(y)=(d B_{z} / d y) y$, and $(d B_{z} / d y)=const.$ 
The magnetic field couples with the Fermion through the Zeeman term 
$\mu B_{z} \bar{\Psi} \sigma_{3} \otimes \gamma_{\tau} \Psi$, 
where $\mu$ is the magnetic moment of the Fermion. 
We note that the Lagrangian is similar to that of 
2D Dirac QED in a background scalar potential, except for the last term 
and the existence of $\sigma_{3}$.  
The axial anomaly in such a system is discussed 
in Ref. \cite{red-nie-sem,ishikawa}.   

Let us calculate the expectation value of the mass density  
\begin{eqnarray}
\left<\rho_{e}({\bf x})\right>
&=& e \left< \bar{\Psi} ({\bf x}) \Psi ({\bf x}) \right>   
\label{j-exp}\\
&=& {\rm Tr}\left[\frac{e \sigma_{3}}{\slad + 
\frac{\Delta}{|p_{\rm F}|} \varepsilon_{x} p_{x} \hat{1} \otimes \gamma_{x}  
- i v_{\rm F} (p_{x} - p_{\rm F}) \sigma_{3} \otimes \hat{1}} \right], 
\nonumber
\end{eqnarray}
where $e$ shows a mass of the quasiparticle.    
It shows an electric charge when we consider a superconductor.  
A hermitian operator $\slad$ in the $y$-direction is defined as 
\begin{equation} 
\slad = \left\{ i \partial_{\tau} \sigma_{3} +  
\mu \frac{d B_{z}}{d y} y \right\} \otimes \gamma_{\tau} 
\nonumber\\
+ 
\frac{\Delta}{|p_{\rm F}|} \varepsilon_{y} p_{y} \hat{1} \otimes \gamma_{y}.   
\end{equation} 
We define $\gamma_{5}$ as $\gamma_{5}=i \gamma_{\tau} \gamma_{y}=-\gamma_{x}$,
and it is anti-commute with $\gamma_{\tau}$ and $\gamma_{y}$, 
therefore, $\gamma_{5}$ is a hermitian matrix and satisfies 
$\{ \gamma_{5}, \slad \}= 0$. 
These facts suggest that 
if an eigenstate $u_{n}$ of $\slad$ with a nonzero 
eigenvalue $\xi_{n}$ ($0<n$) exists 
(i.e. $\slad u_{n} = \xi_{n} u_{n}$), 
$\gamma_{5} u_{n}$ should be another eigenstate with an eigenvalue $-\xi_{n}$.
If zeromodes of $\slad$ exist (i.e. $\slad u_{0} = 0$ and 
$\slad \gamma_{5} u_{0}=0$), they are divided into two groups. 
One of them is $u_{0}^{(+)}=(1/2)(1+\gamma_{5})u_{0}$ 
with an eigenvalue $\gamma_{5}=+1$ and 
another is $u_{0}^{(-)}=(1/2)(1-\gamma_{5})u_{0}$ with 
an eigenvalue $\gamma_{5}=-1$, since $\gamma_{5}^{2}=1$.

Let us research eigenmodes of $\slad$. 
The expectation value of $\slad^{2}$ is 
\begin{eqnarray} 
(u_{n}, \slad^{2} u_{n})&=&|\omega_{c}| 
(n + \frac{1}{2}) 
+ \frac{\omega_{c}}{2} (u_{n}, \gamma_{5} u_{n}),   
\nonumber\\ 
\omega_{c}&=&\mu \frac{d B_{z}}{d y} \frac{2 \Delta}{|p_{\rm F}|} 
\varepsilon_{y}, 
\label{expvalue}
\end{eqnarray}   
where $u_{n}=u_{n}(y - y_{c}(p_{\tau}, \sigma_{3}))$ 
is the eigenfunction of 
the harmonic oscillator with the frequency  
$\omega_{c}$. The oscillator is centered at 
$
y_{c} (p_{\tau},\sigma_{3}) = 
- (d B_{z} / d y)^{-1} (p_{\tau}/\mu) \sigma_{3}. 
$
Eq. (\ref{expvalue}) indicates that 
only zeromodes which belong to $u_{0}^{-}$ ($u_{0}^{+}$) exist 
when $0<\omega_{c}$ ($\omega_{c}<0$).    
It suggests the nonconservation of the vacuum expectation value of 
the axial charge which is defined in the second-quantized formalism as 
\begin{eqnarray}
\left<Q_{5}\right>&=&\left<N_{+} - N_{-}\right>,  
\nonumber\\ 
N_{\pm}&=&\int dp_{y} \hat{u}_{0}^{\dagger (\pm)} \hat{u}_{0}^{(\pm)},  
\end{eqnarray} 
while the classical 1D theory  
${\cal{L}}_{\rm 1D}=\bar{\Psi} \slad \Psi$  
has the axial symmetry $\Psi \rightarrow e^{i \alpha \gamma_{5}} \Psi$,   
i.e. {\it the axial anomaly occurs.} 
Here, $\hat{u}_{0}^{\pm}$ is a second-quantized fermionic field. 
The anomaly comes from the spectral asymmetry 
of zeromodes as same as the discussions 
in Refs.\cite{red-nie-sem,ishikawa,j-l-h}.  
In the free system, the energy spectrum of $u_{0}^{(\pm)}$ is 
$p_{0}= \pm \frac{\Delta}{|p_{\rm F}|} \varepsilon_{y} p_{y}$ 
in Minkowski spacetime, 
and all of the negative energy states are filled   
while all of the positive energy states are empty and $\left<Q_{5}\right>=0$. 
After we turn on the magnetic field adiabatically 
(for a while, we assume $0<\omega_{c}$),  
the energy spectrum of $u_{0}^{(+)}$ is lowered   
and $\left<N_{+}\right>$ decreases (i.e. empty negative energy states arise 
on the spectrum of $u_{0}^{(+)}$), on the other hand,  
the energy spectrum of $u_{0}^{(-)}$ is lifted   
and $\left<N_{-}\right>$ increases (i.e. filled positive energy states arise 
on the spectrum of $u_{0}^{(-)}$), 
therefore $\left<{Q_{5}}\right>$ does {\it not} conserve.  
Finally $\left<N_{+}\right>=0$ and only $u_{0}^{(-)}$ exists.  
The nonzero eigenvalues of $\slad^{2}$ is 
$E_{n}=\omega_{c} (n + 1/2)$, since 
the inner product $(u_{n}, \gamma_{5} u_{n})$ vanishes whenever 
$\slad u_{n} \neq 0$ because of the orthogonal relation between  
the eigenfunctions of the hermitian operator.

Next, we consider the eigenvalue problem of a 2D operator
\begin{equation}
\slaD = \slad + 
\frac{\Delta}{|p_{\rm F}|} \varepsilon_{x} p_{x} \hat{1} \otimes \gamma_{x} 
= 
\slad - 
\frac{\Delta}{|p_{\rm F}|} \varepsilon_{x} p_{x} \hat{1} \otimes \gamma_{5}.
\end{equation}
Let 
\begin{equation}
\varphi_{n}= (\alpha_{n} u_{n} + \beta_{n} \gamma_{5} u_{n}) 
e^{i p_{x} x}
\end{equation}
stands for an eigenfunction. 
We use a representation for the $n$-th level such as ($\nearrow$)
\begin{equation} 
\slad= 
\left( \begin{array}{cc} 
\xi_{n} & 0  \\
0 & - \xi_{n}  
\end{array} \right) , 
u_{n}=
\left( \begin{array}{c} 
1 \\
0
\end{array} \right) ,  
\gamma_{5}=  
\left( \begin{array}{cc} 
0 & 1 \\
1 & 0 
\end{array} \right), 
\end{equation} 
where, 
$$
\xi_{n}=
\left\{\begin{array}{cl}
\sqrt{|\omega_{c}| (n + \frac{1}{2})} 
& (n = 1,2,\cdot\cdot\cdot), 
\\
0 & (n=0). 
\end{array}\right.
$$
Therefore, the eigenvalue equation is written as  
\begin{equation}
\left(\begin{array}{cc}
\xi_{n} & -\frac{\Delta}{|p_{\rm F}|} \varepsilon_{x} p_{x} \\ 
-\frac{\Delta}{|p_{\rm F}|} \varepsilon_{x} p_{x} & -\xi_{n}  
\end{array} \right)
\left(\begin{array}{c}
\alpha_{n} \\ 
\beta_{n} 
\end{array} \right)
=\zeta_{n}
\left(\begin{array}{c}
\alpha_{n} \\ 
\beta_{n} 
\end{array} \right).   
\end{equation} 
There are two eigenstates for an oscillator in the $n (\neq 0)$-th level  
written as  
\begin{eqnarray}
&\zeta_{n}^{(\pm)}(p_{x})&
=\pm \sqrt{\xi_{n}^{2} + 
\frac{\Delta^{2}}{p_{\rm F}^{2}} \varepsilon_{x}^{2} p_{x}^{2}},  
\\
&\left(\begin{array}{c}
\alpha_{n}^{+} \\
\beta_{n}^{+} 
\end{array} \right)&  
=\frac{1}{C_{+}} 
\left(\begin{array}{c} 
\zeta_{n}^{(+)} + \xi_{n}  \\
-\frac{\Delta}{|p_{\rm F}|} \varepsilon_{x} p_{x} 
\end{array} \right) ,  
\nonumber\\
&\left(\begin{array}{c}
\alpha_{n}^{-} \\
\beta_{n}^{-} 
\end{array} \right)&
=\frac{1}{C_{-}} 
\left(\begin{array}{c} 
\frac{\Delta}{|p_{\rm F}|} \varepsilon_{x} p_{x} \\ 
-\zeta_{n}^{(-)} + \xi_{n} 
\end{array} \right), 
\nonumber
\end{eqnarray}
where $C_{\pm}$ are normalization constants, but for $n=0$, there is only 
one eigenstate   
\begin{eqnarray} 
\zeta_{0}(p_{x})&=& \frac{\omega_{c}}{|\omega_{c}|} 
\frac{\Delta}{|p_{\rm F}|} \varepsilon_{x} p_{x} 
,  
\nonumber\\
\left(\begin{array}{c} 
\alpha_{0} \\
\beta_{0} 
\end{array} \right) 
&=& \frac{1}{2}
\left(\begin{array}{c} 
1 \\
- \omega_{c}/|\omega_{c}| 
\end{array} \right), 
\end{eqnarray} 
because the solution should satisfy 
$\gamma_{5} \varphi_{0} = - (\omega_{c}/|\omega_{c}|)\varphi_{0}$. 
This condition comes from the axial anomaly in the $y$-direction.  

Finally, we show the accumulation of 
the mass density from Eq. (\ref{j-exp}),  
which is derived as
\end{multicols}
\begin{eqnarray} 
\left<\rho_{e}({\bf x})\right>&=& {\rm Tr} 
\left[ 
\frac{e \sigma_{3}} 
{\slaD - i v_{\rm F}(p_{x} - p_{\rm F})\sigma_{3} \otimes \hat{1}} 
\right] 
=\sum_{n}
\int_{-\infty}^{\infty} \frac{d p_{\tau}}{2 \pi} 
\int \frac{d p_{x}}{2 \pi} {\rm tr} \left[ 
\frac{e \sigma_{3}   
|u_{n}(y - y_{0}(p_{\tau}, \sigma_{3}))|^{2}}  
{\zeta_{n}(p_{x})  - 
i v_{\rm F} (p_{x} - p_{\rm F}) \sigma_{3}} \right] 
\nonumber\\
&=& \frac{-e \mu}{2 \pi}\frac{d B_{z}}{d y} \sum_{n \neq 0}    
\int \frac{d p_{x}}{2 \pi} {\rm tr} \left[ 
\frac{1}  
{\zeta_{n}^{(+)}(p_{x})  - 
i v_{\rm F} (p_{x} - p_{\rm F}) \sigma_{3}} 
+
\frac{1}  
{\zeta_{n}^{(-)}(p_{x})  - 
i v_{\rm F} (p_{x} - p_{\rm F}) \sigma_{3}} 
\right]
\nonumber\\
&&
- \frac{e \mu}{2 \pi} \frac{d B_{z}}{d y}    
\int \frac{d p_{x}}{2 \pi} {\rm tr} \left[ 
\frac{1}  
{\zeta_{0}(p_{x}) - 
i v_{\rm F} (p_{x} - p_{\rm F})\sigma_{3}} \right]
=- {\rm sgn}(\varepsilon_{x} \varepsilon_{y}) 
e \mu N_{\rm 1D}(0) 
\frac{d B_{z}}{d y},    
\end{eqnarray}
\begin{multicols}{2}
where the symbol $tr$ means a trace on the real spin, 
and we use the normal-orthogonal relation $\int dy |u_{n}|^{2}=1$.     
$\int \frac{d p_{x}}{2 \pi}
=\int_{p_{\rm F}-\Lambda}^{p_{\rm F}+\Lambda}
\frac{d p_{x}}{2 \pi}$ 
, and $\Lambda$ is a momentum cutoff.  
We assume a relation $|\Delta|<<\Lambda^{2} / 2 m<<\epsilon_{F}$.    
$N_{\rm 1D}(0)=(2 \pi v_{\rm F})^{-1}$ is the density of state at 
the Fermi surface in 1D.
All of the $n \neq 0$ parts are canceled out because of the 
co-existence of the eigenvalues $\zeta_{n}^{(+)}$ and $\zeta_{n}^{(-)}$.     
{\it Only the $n=0$ part survives because of the axial anomaly in the 
$y$-direction}. 

We can define a chiral transformation in the $x$-direction 
such as $\psi(\pm p_{\rm F}, {\bf x}) \rightarrow 
e^{\pm i \alpha} \psi(\pm p_{\rm F}, {\bf x})$, therefore 
$\Psi \rightarrow e^{i \alpha} \Psi, \bar{\Psi} \rightarrow 
\bar{\Psi}e^{i \alpha}$. The expectation value of 
the corresponding current density which is perpendicular to $d B_{z} / dy$ is  
\begin{equation} 
\left< j_{x}^{chi} ({\bf x}) \right> 
=e v_{\rm F} \left< \bar{\Psi}({\bf x})\Psi({\bf x}) \right>  
=- {\rm sgn}(\varepsilon_{x} \varepsilon_{y}) \frac{e \mu}{2 \pi}  
\frac{d B_{z}}{d y},    
\end{equation} 
and we call it a chiral Hall current density. 

These two effects are related to the winding number of   
the gap Eq. (\ref{c-p-w})\cite{vol-g-i},   
\begin{eqnarray}
{\rm sgn}(\varepsilon_{x} \varepsilon_{y})
&=&\int \frac{d^{2} p}{16 \pi} 
tr[\hat{{\bf g}} \cdot 
({\bf \nabla}\hat{{\bf g}} \times {\bf \nabla} \hat{{\bf g}})]  
, 
\\  
{\bf g}({\bf {p}})&=& 
\left(\begin{array}{c} 
{\rm Re}[\Delta({\bf {p}})(-i \sigma_{2})]  \\ 
-{\rm Im}[\Delta({\bf {p}})(-i \sigma_{2})] \\
({\bf p}^{2} / 2 m) - \epsilon_{\rm F}
\end{array} \right),  
\nonumber
\end{eqnarray}
where ${\bf \nabla}=\partial / \partial {\bf p}$. 
It suggests that these effects occur even if $\varepsilon_{x}$ and/or  
$\varepsilon_{y}$ are infinitesimally small, and that these effects 
come from the P- and T-violation of the gap.

The accumulated mass density and the chiral Hall current density  
exist in the bulk region of the superfluid.  
In the superconductors, the Meissner effect occurs 
and the magnetic field cannot penetrate into the bulk, therefore  
the accumulated charge density and the chiral Hall 
current density would exist near the edge 
of the superconductors\cite{note} and also around the vortex core. 
As we mentioned before, 
our discussion could be applicable to Sr$_{2}$RuO$_{4}$ near the 
second superconducting phase transition point, 
some quasi-1D organic superconductors
and the FQH state 
at $\nu = 5 / 2$ LL filling factor
by varying the parameters $\varepsilon_{x}$ and $\varepsilon_{y}$.  
Recently, the vortex in chiral superconductors has 
been discussed \cite{goryo}, and such a vortex has a fractional charge and 
a fractional angular momentum.   
Interesting phenomena related to these fractional quantum numbers 
and the present effects are expected to occur 
around the vortex core.

The axial anomaly also causes the spin quantum Hall 
effect (SQHE) in the chiral $d$-wave ($d_{x^{2}-y^{2}} + i d_{xy}$-wave) 
superconductors\cite{fisher-sqhe}.  
The low energy quasiparticles in a magnetic field 
with a constant gradient can be mapped onto the massive Dirac Fermion in 
a constant electric field, and the spin rotation around the $z$-axis 
for the quasiparticle   
corresponds to the $U(1)$ transformation for the Dirac Fermion. 
Therefore, according to the 
discussions in Ref.\cite{red-nie-sem,ishikawa}, 
we can see that the axial anomaly causes the quantized spin Hall current, 
which is perpendicular to the gradient of the 
magnetic field. 

SQHE has been pointed out by Volovik and Yakovenko 
in superfluid $^{3}$He-A film, 
which is the chiral $p$-wave superfluid\cite{volovik-1,vol-yak}.  
They have described the effect by the Chern-Simons term.     
It has been clarified the relation between the axial anomaly 
and the Chern-Simons term in 2D Dirac QED\cite{ishikawa}.  
Therefore, SQHE in $^{3}$He-A could be related 
to the axial anomaly. According to Ref.\cite{vol-yak}, 
SQHE also occurs at the edge or around the vortex core of
 the superconducting Sr$_{2}$RuO$_{4}$\cite{cpsc} by 
a magnetic field in the basal plane. 

The author thanks K. Ishikawa and N. Maeda for useful discussions and 
encouragement. 
This work was partially supported by the special Grant-in-Aid for 
Promotion of Education and Science in Hokkaido University provided by 
the Ministry of Education, Science, Sports, and Culture, the Grant-in-Aid 
for Scientific Research on Priority area 
(Physics of CP violation)
(Grant No. 10140201), and the Grant-in-Aid for International Science Research 
(Joint Research 10044043) from the 
Ministry of Education, Science, Sports and Culture, Japan.

\end{multicols}


\begin{references}
\bibitem{note} Note that the axial anomaly discussed here  
is in the $y$-direction, which is perpendicular to the edge, 
therefore it is different from that of the 
chiral edge mode\cite{edge}.
\end{references}
\end{document}